\documentclass[aps,pra,groupedaddress,superscriptaddress,twocolumn,showpacs]{
revtex4-1}

\usepackage[utf8x]{inputenc}
\usepackage{color}
\usepackage{bbm} 

\usepackage{amsfonts,amsmath,amssymb,stmaryrd}

\usepackage{graphicx}

\usepackage{subfigure}  

\usepackage{hyperref}

\usepackage{epsfig}

\usepackage{mathrsfs}

\usepackage{verbatim}

\usepackage{ulem}	

\renewcommand{\L}{\left(}
\newcommand{\R}{\right)}
\newcommand{\bra}[1]{\langle#1|}
\newcommand{\ket}[1]{|#1\rangle}

\newcommand{\av}[1]{\left\langle #1 \right\rangle}

\newcommand{\Jastrow}{\prod_{j<k}(z_j-z_k)}
\newcommand{\CF}{\mathrm{CF}}

\newcommand{\Ch}{\mathrm{Ch}}
\newcommand{\lB}{\ell_\mathrm{B}}
\newcommand{\LChB}{\mathrm{LChB}}
\newcommand{\HChB}{\mathrm{HChB}}
\newcommand{\Edge}{\mathrm{Edge}}

\renewcommand{\P}{\hat{\mathcal{P}}}
\newcommand{\F}{\mathcal{F}}
\newcommand{\LLL}{\text{LLL}}

\newcommand{\LN}{\text{LN}}
\newcommand{\qh}{\text{qh}}
\newcommand{\eff}{\text{eff}}
\renewcommand{\max}{\mathrm{max}}

\newcommand{\e}{\mathrm{e}}
\renewcommand{\d}{\mathrm{d}}
\renewcommand{\l}{\ell}

\newcommand{\dt}{\partial_t}

\renewcommand{\H}{\hat{\mathcal{H}}}
\renewcommand{\c}{\hat{c}}
\renewcommand{\a}{\hat{a}}

\newcommand{\cd}{\hat{c}^\dagger}
\newcommand{\ad}{\hat{a}^\dagger}
\newcommand{\bd}{\hat{b}^\dagger}
\renewcommand{\b}{\hat{b}}

\newcommand{\hc}{\text{h.c.}}

\newcommand{\psd}{\hat{\psi}^\dagger}
\newcommand{\ps}{\hat{\psi}}

\usepackage{array}

\usepackage{bm}	
\renewcommand{\vec}[1]{\bm{#1}}

\usepackage{cancel,ifthen}
\newcommand{\cmnt}[2][NoInPuT]{\ifthenelse{\equal{#1}{NoInPuT}}{}{{\color{red}\sout{#1}}} {\color{blue} #2}}

\begin{document}
\normalem	

\title{Growing quantum states with topological order}

\author{Fabian Letscher}
\affiliation{Department of Physics and Research Center OPTIMAS, University of 
Kaiserslautern, Germany}

\author{Fabian Grusdt}
\affiliation{Department of Physics and Research Center OPTIMAS, University of 
Kaiserslautern, Germany}
\affiliation{Graduate School Materials Science in Mainz, Gottlieb-Daimler-Strasse 47, 
67663 Kaiserslautern, Germany}

\author{Michael Fleischhauer}
\affiliation{Department of Physics and Research Center OPTIMAS, University of 
Kaiserslautern, Germany}

\begin{abstract}
We discuss a protocol for growing states with topological order in 
interacting many-body systems using a sequence of flux quanta and particle 
insertion. We first consider a simple toy model, the superlattice Bose Hubbard 
model, to explain all required ingredients. Our protocol is then applied to 
fractional 
quantum Hall systems in both, continuum and lattice. We investigate in 
particular
how the fidelity, with which a topologically ordered state can be grown, scales 
with increasing particle number $N$. For small systems exact diagonalization methods 
are used. To treat large systems with many particles, we introduce an effective model 
based on the composite fermion description of the fractional quantum Hall effect. 
This model also allows to take into account the effects of dispersive bands and edges 
in the system, which will be discussed in detail. 
\end{abstract}

\date{\today}

\maketitle
\section{Introduction}
Since the discovery of the integer quantum Hall effect (IQHE) in 1980 
\cite{Klitzing1980} and two years later the fractional quantum Hall effect (FQHE) 
\cite{Tsui1982}, the interest in states exhibiting topological order has increased 
tremendously. Due to their robustness against local disorder, these states are for 
instance well suited for metrology \cite{Klitzing1980}. Another interesting aspect of 
such interacting many-body systems is that they host states with anyonic 
excitations \cite{Arovas1984,Halperin1984,Moore1991}. In this context, non-Abelian 
anyons are particularly exciting because they can be employed to build a topological 
quantum computer \cite{Nayak2008}, if they can be coherently manipulated.

For many years, solid state systems with electrons appeared to be the only candidates
to realize exotic many-body states, e.g. in the FQHE. However, their small 
intrinsic length scales make coherent control difficult. On the other hand, the 
rapid development of ultracold gases and photonic systems is a promising route to 
implement various Hamiltonians known from the solid state community, with 
full coherent control. With current state of the art technology, noninteracting 
topological states have been observed in ultracold gases \cite{Aidelsburger2013, 
Miyake2013, Jotzu2014, Aidelsburger2014} as well as photonic systems 
\cite{Hafezi2013, Rechtsman2013}. 

\begin{figure}[t]
\centering
\epsfig{file=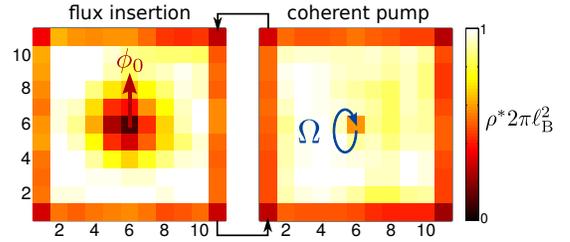, width=0.4\textwidth}
\caption{(Color online) Simulation of the effective composite fermion (CF) lattice 
model. The figure shows the CF density $\rho^*$ of a lattice with 
$11 \times 11$ sites and effective flux per plaquette $\alpha^* = 0.1$. The basic 
ingredients of the protocol are ($i$) flux insertion, which creates a CF quasi-hole 
excitation and ($ii$) a coherent CF pump, which refills the hole excitation. 
Additionally, we include a quasi-hole trapping potential ($g_\mathrm{h} = 1J^*$) 
and absorbing 
boundaries ($\gamma_\Edge = 0.0025J^*$) to account for the effects of dispersive 
bands and 
edge states. }
\label{fig:LatticeDensity_Protocol}
\end{figure}

Unlike in the solid state context, efficient cooling mechanisms below the many-body 
gap are still lacking in systems of interacting atoms and photons, which makes the 
preparation of topological states challenging. In this paper we develop an 
alternative dynamical scheme for the efficient preparation of highly correlated 
ground states with topological order. It can be implemented with ultracold atoms or 
photons and exploits the coherent control techniques available in these systems.

To solve the cooling problem, different approaches were discussed previously. In 
\cite{Umucalilar2012} it was suggested to pump 
the sites of a dissipative lattice system coherently. This scheme relies on a 
direct $N$-photon transition and thus works only for small systems. Moreover, the 
state 
prepared is in a superposition of different particle numbers. Another proposal 
\cite{Kapit2014} suggested to rapidly refill hole excitations due to local loss in a 
lattice system. This protocol stabilizes the ground state dynamically. 

In reference \cite{Grusdt2014a}, we proposed a scheme to grow the highly correlated 
Laughlin states. Here, we discuss the general ingredients of the protocol, which may 
also be used to grow other topological ordered states in interacting many-body 
systems. 
To do so, we consider a simple toy model explaining all necessary ingredients. The 
main ideas of the protocol can be 
summarized as follows. \textit{In the first step, a topologically protected and 
quantized Thouless pump \cite{Thouless1982} is used to create a localized hole 
excitation in the system. In the second step the hole is refilled using a coherent 
pump. Local repulsive interactions between the particles are employed to provide a 
blockade mechanism such that only a single particle is inserted.} In the 
case of the 
FQHE, we analyze the performance of the protocol in detail and explain how the 
fidelity scales with the particle number $N$. In order to treat large fractional 
Chern insulator systems with many particles, we introduce an effective model based on 
the composite fermion (CF) description of the FQHE. Within this model, we are able to
demonstrate, that the growing scheme still works despite the presence of gapless 
edge states and dispersive bands in large lattice systems. As shown in Fig. 
\ref{fig:LatticeDensity_Protocol}, we reach a homogeneous density in the bulk with 
an average CF magnetic filling factor $\nu^* \simeq 0.9$, close to the optimal value 
$\nu^* = 
1$.

The paper is structured as follows. We start by describing the growing scheme using 
a simple toy model in section \ref{sec:GrowingScheme}. In section 
\ref{sec:Continuum}, we discuss the protocol in a FQH system. Moreover, we discuss 
the performance of the protocol. The last section \ref{sec:Lattice} considers 
fractional Chern insulators. There, we discuss edge and band dispersion effects based 
on our effective CF model. 

\section{Topological growing scheme}\label{sec:GrowingScheme}

In this section, we discuss the basic idea how states with topological order 
can be grown for the one dimensional superlattice Bose Hubbard model (SLBHM) as 
a toy model. This sets the stage for the following discussions of two dimensional 
systems with topological order.

\subsection{Toy model}\label{subsec:GrowingSchemeToyModel}
\begin{figure}[t]
\centering
\epsfig{file=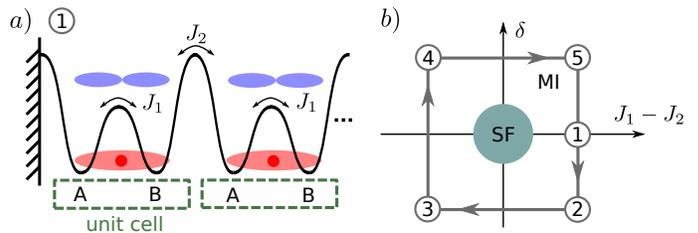, width=0.5\textwidth}
\caption{(Color online) (a) Semi-infinite chain of coupled dimers, consisting of 
two sites $A,B$. Initially, the MI ground state is prepared with filling $\rho = 
1/2$. (b) Schematic phase diagram of the SLBHM in the regime of intermediate 
interactions $U$. Following adiabatically the loop 1-5 corresponds to a Thouless 
pump: Encircling the SF region, all particles in the MI are shifted one dimer to 
the right. }
\label{fig:SLBHMandPhaseDiagram}
\end{figure}

The SLBHM can be used to implement a quantized pump \cite{Berg2011, Thouless1982}, 
which can be related to a nontrivial topological invariant. The Hamiltonian of the 
SLBHM is
\begin{multline}
 \H = -J_1 \sum_j \ad_j \b_j + \hc - J_2 \sum_j \ad_j \b_{j+1} + \hc 
\nonumber \\
+ \delta \sum_j \bd_j \b_j 
+ U/2 \sum_j \L \ad_j \ad_j \a_j \a_j + \bd_j \bd_j \b_j \b_j \R,
\end{multline}
where $\a_j (\b_j)$ annihilates a boson on lattice site $A$ ($B$) in the 
$j$th unit cell (see Fig. \ref{fig:SLBHMandPhaseDiagram}a). Note, that we consider 
here a semi-infinite system with open boundary on the left. The hopping elements are 
denoted by $J_1, J_2$, while $\delta$ is an onsite potential shift acting on lattice 
sites $B$. 
Furthermore, we include onsite interactions $U$ for more than one particle per site. 

In the case of vanishing $\delta$ and disregarding interactions, the different 
hopping amplitudes $J_1 \neq J_2$ 
determine the dimerization of the two sites $A$ and $B$ in the unit cell. This leads 
to a two-band model with a band gap $E_\mathrm{gap}$ and bandwidth $\Delta E$ 
determined by the ratio $J_2/J_1$. In the limit $J_2/J_1 \rightarrow 0$ the 
flatness ratio $E_\mathrm{gap}/\Delta E$ tends to infinity and the two bands become 
nondispersive. At filling $\rho = 1/2$ and with interactions, the ground state of 
the system 
is a Mott insulator (MI) with many-body gap $\Delta$ or a superfluid (SF) depending 
on the specific parameters $J_2/J_1$ and $U$ \cite{Buonsante2004, 
Buonsante2004a, Buonsante2005, Muth2008}. Specifically, for large 
interactions $U$ the model can be mapped to free fermions. The resulting ground state 
at $\rho=1/2$ is incompressible whenever $J_1 \neq J_2$ or $|\delta| \neq 0$. For 
finite interaction $U$, the SF region is extended in parameter space $(J_1-J_2, 
\delta)$ as depicted in Fig. 
\ref{fig:SLBHMandPhaseDiagram}b. 

\subsection{Protocol}
In the following, we illustrate the main steps of our growing scheme by showing how 
the MI ground state of this system can be grown. We note that there are more 
practical ways to prepare the ground state, of course, but the present protocol can 
be generalized to more complex systems, including states in 
the FQHE presented in section \ref{sec:Continuum} and \ref{sec:Lattice}.  

We will start to discuss the protocol in the nondispersive limit $J_2/J_1 \rightarrow 
0$, where hole excitations will remain localized. The 
effect of dispersive bands in the case of finite $J_1 / J_2$ will be considered 
later. Let us assume, that the $N$ particle MI ground state of the 
SLBHM with filling $\rho = 1/2$ is already prepared in a finite region of the 
lattice. We now show how the state with $N+1$ particles can be grown. 

\subsubsection{Topological pump}
First, a topological Thouless pump \cite{Thouless1982} is used to create a hole 
excitation localized at the edge of the system. This process is related to 
Laughlin's argument of flux insertion \cite{Laughlin1981} in the quantum Hall effect 
(QHE): Inserting one flux quantum $\phi_0$ leads to a quantized particle transport, 
which is the origin of the quantized Hall conductance 
$\sigma_{xy}$. In the SLBHM, the process of flux insertion is shown in Fig. 
\ref{fig:SLBHMandPhaseDiagram}b. By adiabatically changing the parameters 
$J_1-J_2$ and $\delta$ in time we encircle the critical SF region. After a full 
cycle of period $T_\phi$ all particles are shifted one dimer to the right, leaving a 
hole excitation on the left side of the system. The possibility to construct such a 
cyclic pump is deeply related to the underlying topological order of the model. 

\subsubsection{Coherent pump}

\begin{figure}[t]
\centering
\epsfig{file=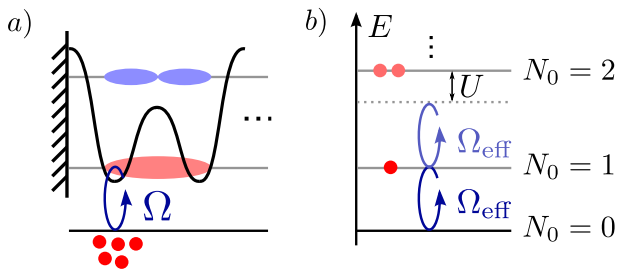, width=0.45\textwidth}
\caption{(Color online) (a) Coherent pump couples a reservoir of bosons to the 
left-most dimer. (b) Due to a blockade mechanism, only a single 
particle is pumped into the system. $\Omega_\eff$ is the many-body Rabi-frequency, 
reduced by a Franck-Condon factor and $N_0$ the number of bosons in the left-most 
dimer.}
\label{fig:SLBHMCoherentPumpBlockade}
\end{figure}

In the following step, we replenish the hole excitation by a single boson. To this 
end, we coherently couple a reservoir of particles to the first lattice site of the 
system as shown in Fig. \ref{fig:SLBHMCoherentPumpBlockade}a. The coherent pump can 
be described by the Hamiltonian
\begin{equation}
\H_\Omega = \Omega \ad_0 \e^{-i \omega t} + \hc ,
\label{eq:Hcoherent}
\end{equation}
where $\Omega$ is the Rabi- and $\omega$ the driving frequency. The driving frequency 
$\omega$ is chosen to be resonant with the lowest band. Hence, if $|\Omega|$ 
is much smaller than the single-particle band gap $E_\mathrm{gap}$, particles can 
only be added in the lowest band. The corresponding Rabi-frequency $|\Omega_{\rm 
eff}| < |\Omega|$ is then however reduced by a Franck-Condon factor.

In general, the coherent pump \eqref{eq:Hcoherent} creates a coherent superposition 
state of $N_0$ bosons in the left-most Wannier orbital of the lowest band. To ensure 
that, at most, one boson is added, we employ a blockade mechanism, see Fig. 
\ref{fig:SLBHMCoherentPumpBlockade}b. For sufficiently strong interactions and weak 
enough driving,
\begin{equation}
\label{eq:BlockadeCondition}
 U \gg |\Omega_{\rm eff}|
\end{equation}
the transition to the $N_0=2$ particle state is detuned by the interaction energy 
$U$.

For $|\Omega_{\rm eff}| \ll \Delta$, at maximum one boson can be 
added into the lowest Bloch band. To replenish the hole excitation by precisely 
one boson at a time, one can use a $\pi$-pulse of duration $T_\Omega = 
\pi/2\Omega_\eff$. Although the $\pi$-pulse is less robust to errors than 
e.g. an adiabatic sweep, we choose it because of its speed. This, we believe, is 
crucial to overcome linear losses present in realistic systems.

Starting from complete vacuum, the sequence described above can be repeated to grow 
the MI ground state with $N$ particles. Next, we discuss extensions of the protocol, 
required when the bands are dispersive or the system is finite.

\subsubsection{Dispersive bands \& finite systems}

\begin{figure}[t]
\centering
\epsfig{file=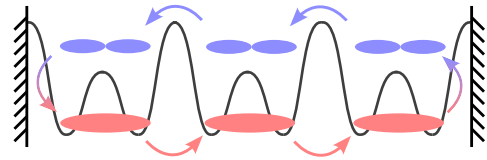, width=0.35\textwidth}
\caption{(Color online) Due to the process of flux insertion in a finite system, the 
higher band of the SLBHM will be occupied. While a state at the right edge of the 
system will be transferred into the upper band, a state from the left upper band, 
will be transferred to the lower band. } 
\label{fig:SLBHMEdgeEffects}
\end{figure}

Dispersive bands $J_1 / J_2 \neq 0$ lead to intrinsic dynamics of the particles and 
thus of the hole excitations. The bandwidth $\Delta E$ of the lowest band sets the 
timescale for the dispersion. Firstly, the localized quasi-hole excitation created 
using the topological pump, disperses into the bulk of the system. Since the 
coherent pump couples locally to the left-most dimer, the hole cannot be refilled 
efficiently with a single boson. More importantly, the number fluctuations $\Delta 
N(t) = \left[ \big\langle \hat N^2 \big\rangle - \big\langle \hat N\big\rangle^2 
\right]/\big\langle \hat N\big\rangle$ will increase. To prevent the 
dynamics of the quasi-hole excitation, a quasi-hole trapping potential
\begin{equation}
 \label{eq:ToyModelHoleTrapping}
 \H_\mathrm{qh} = g_\mathrm{h} \L \ad_0 \a_0 + \bd_0 \b_0 \R
\end{equation}
acting on the left-most dimer can be used. The strength of the trap $g_\mathrm{h}$ 
should 
be weak enough not to destroy the topological order, but strong enough to trap the 
quasi-hole excitation. Concretely we require 
\begin{equation}
\Delta E \lesssim g_{\rm h} \ll \Delta.
\end{equation}

Secondly, the diffusion of bulk particles from the right edge of the system into 
vacuum lets the MI melt. Vice-versa, the diffusion of holes from the vacuum into the 
bulk at the right edge of the system makes the state compressible.  For fast pump 
rates of the protocol, this effect can be neglected. However, if the bandwidth 
$\Delta E$ is the dominant contribution, we suggest to use a harmonic trapping 
potential. The 
potential should be weak enough to avoid localization effects (see 
\cite{Kolovsky2014}). Therefore, in local density approximation, the bulk of the 
system will remain incompressible. 

In finite systems with open boundaries, the flux insertion, as illustrated in Fig. 
\ref{fig:SLBHMEdgeEffects}, connects states from the lower and the upper band 
at the edges. As long as the upper band is empty, the 
topological pump creates a hole excitation in the lower band. However, once the 
right boundary is reached, particles will be excited to the upper band. To avoid 
the high energy excitations in the protocol, sufficiently strong losses 
$\gamma_\Edge$ localized at the right boundary of the system can be introduced.

\section{Fractional Quantum Hall States}\label{sec:Continuum}

We now apply the growing scheme to fractional quantum Hall (FQH) systems. In this 
case, the bands -- i.e. Landau levels (LLs) -- are nondispersive and we assume an 
infinite system. 

\subsection{Model}\label{subsec:ContinuumModel}
We consider a FQH model of bosons in the two dimensional disc geometry. The magnetic 
field can be implemented using e.g. artificial gauge fields. Moreover, we consider a 
contact interaction between the particles with strength $g_0$. In second quantized 
form, the Hamiltonian reads
\begin{multline}
 \label{eq:FQHE}
 \H = \int \d^2 \vec r \ \psd(\vec r) \frac{1}{2M} \L \vec p - \vec A \R^2 \ps(\vec 
r) \\
+ \frac{1}{2}g_0 \int \d^2 \vec r \ \psd(\vec r)\psd(\vec r)\ps(\vec r)\ps(\vec r),
\end{multline}
where $\psd(\vec r)$ creates a boson at the position $\vec r$. The first term in eqn. 
(\ref{eq:FQHE}) includes the magnetic field in minimal coupling using the vector 
potential $\vec A = B/2 (-y,x,0)$. In this symmetric gauge, the total angular 
momentum $L_z$ is a conserved quantity, $[\H,L_z]=0$. 

For later purposes we express the field operators $\ps (\vec r)$ in terms of the 
bosonic operators $\b_{n,\l}$ which create a particle in the orbital of the $n$th 
LL with angular momentum $\l$. Here, $n$ and $\l$ are integers with 
$n,\l + n \geq 0$. We obtain
\begin{equation}
 \ps(\vec r) = \sum_{n,\l} \eta_{n,\l}(\vec r) \b_{n,\l},
\end{equation}
where $\eta_{n,\l}(\vec r)$ are the single particle wavefunctions, see e.g. 
\cite{Prange1987, Jain2007}. 

\subsection{Laughlin States and Excitations - CF 
Picture}\label{subsec:ContinuumStates}

It was shown in \cite{Wilkin1998, Wilkin2000,Cooper1999, Regnault2003, 
Regnault2004}, that the Laughlin (LN) 
wavefunction at magnetic filling $\nu = 1/2$ is the exact zero 
energy ground state of the bosonic FQH Hamiltonian (\ref{eq:FQHE}). The filling 
factor $\nu = N/N_\phi$ is defined as the ratio between particle number $N$ and 
number of flux quanta $N_\phi$. In terms of the complex coordinates $z_j = x_j - i 
y_j$ of the $j$th particle, the LN wavefunction is
\begin{equation}
  \label{eq:LNwavefunction}
 \ket{\LN,N} \ \hat= \ \Psi_\LN = \mathcal{N}_{\rm LN}
	  \Jastrow^2 e^{ - \sum_j |z_j|^2 / 4 \lB^2 },
\end{equation}
where $\mathcal{N}_{\rm LN}$ is a normalization constant. The magnetic length $\lB = 
\sqrt{\hbar/B}$ depends only on the magnetic field $B$. 

The zero energy excitations of the LN wavefunction are quasi-holes described by the 
wavefunction $\Psi_{\rm qh} = \mathcal{N}_{\rm qh} \prod_j z_j \Psi_{\rm LN}$. $m$ 
quasi-holes, located in the center, are described by the $m$ quasi-hole wavefunction
\begin{equation}
 \label{eq:qhwavefunction}
 \ket{m\qh,N} \ \hat= \ \Psi_{m\qh} = \mathcal{N}_{m \rm qh} \Big( \prod_j z_j 
\Big)^m \Psi_\LN.
\end{equation}
The LN state $\ket{\LN,N}$ and the quasi-hole state $\ket{m\qh,N}$ carry different 
total angular momentum $L_z$, 
\begin{align}
\label{eq:AngularMomentumLNstate}
 L_z \big( \ket{\LN,N} \big) & = N(N-1) \\
\label{eq:AngularMomentumQuasiHoleState}
 L_z \big( \ket{m\qh,N} \big) & = N(N-1) + m N.
\end{align}

\begin{figure*}[t]
\centering
\epsfig{file=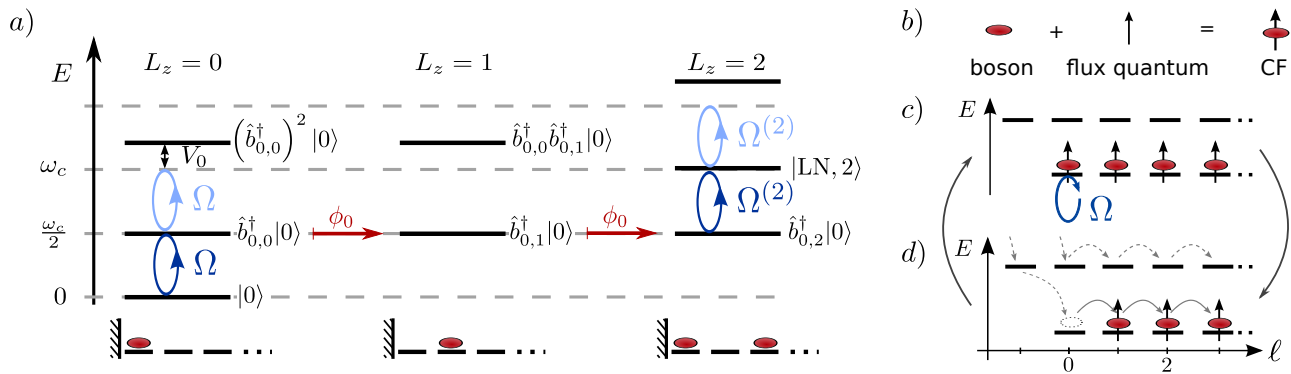, width=0.95\textwidth}
\caption{(Color online) (a) Growing scheme for the $\nu = 1/2$ LN state 
in the continuum. The many-body eigenstates of the FQH Hamiltonian \eqref{eq:FQHE}, 
between which transitions are driven, are shown as a function of the total conserved 
angular momentum $L_z$. Starting from the vacuum $\ket{0}$, the LN state with $N=2$ 
bosons is grown. After adding one boson in the central orbital, two flux quanta are 
adiabatically inserted and the resulting hole excitation is refilled by a second 
boson. This can be interpreted as replenishment of the system by a composite fermion, 
hence keeping intact the topological order of the system. The lowest line 
shows the schematics of the prepared states after every step. (Horizontal lines 
denote orbitals; This picture is valid in the thin-torus limit or after using the 
patterns as partitions determining certain Jack polynomials \cite{Bernevig2008}.)
(b) Flux attachment: A single boson together with one flux 
quantum form a CF. (c),(d) Level spectrum in terms of CFs. (c) Coherent pump process 
with Rabi-frequency $\Omega$. The lowest CF-LL is filled, corresponding to CF 
filling factor $\nu^* = 1$. (d) Flux insertion process increases angular momentum 
$\l$ by one. } 
\label{fig:ContinuumProtocol}
\end{figure*}

Both wavefunctions can be understood in the more general framework of composite 
fermions (CFs) \cite{Zhang1989, Read1989, Jain1989, Jain2007}. For the $\nu = 1/2$ 
case, the LN wavefunction \eqref{eq:LNwavefunction} can be decomposed into
\begin{equation}
 \Psi_\LN = \mathcal{N}_{\rm LN} \Jastrow \Phi_\CF^{(\nu^* = 1)}.
\end{equation}
Besides the Jastrow factor $\Jastrow$ attaching one flux quantum to each boson (see
Fig. \ref{fig:ContinuumProtocol}b), a fermionic wavefunction $\Phi_\CF^{(\nu^* = 
1)}$ 
appears. The Jastrow factor screens the interactions between the particles and 
leads to a reduced magnetic field $B^* = B/2$ seen by the CFs. As in the IQHE, their 
wavefunction is given by a Slater determinant of single particle orbitals filling 
the lowest CF-LL $\nu^* = 1$, i.e. $\Phi_\CF^{(\nu^* = 1)} = \Jastrow \exp \L - 
\sum_j |z_j|^2 / 4 \lB^2 \R$. The FQHE can be interpreted as an IQHE of 
noninteracting CFs in a reduced magnetic field $B^*$. The CF picture in the 
context of 
the FQHE is powerful in describing all fractions in the Jain sequence 
\cite{Jain1989} and moreover describes the quasi-hole excitations and their counting 
correctly. However, the CF theory does not reproduce the correct Laughlin gap 
$\Delta_\LN$. Here, a microscopic theory of the full interacting many-body problem is 
necessary. Also in order to explain other than the main sequence of FQH states, 
interactions between CFs need to be taken into account.

\subsection{Protocol}\label{subsec:ContinuumProtocol}

\subsubsection{Topological pump -- flux insertion}

In the toy model \ref{subsec:GrowingSchemeToyModel} we used a topological Thouless 
pump to shift the particles to the right. In the context of quantum Hall physics, 
this corresponds to Laughlin's argument of flux insertion \cite{Laughlin1981}, which 
was used to explain the quantization of the Hall conductivity. The idea is to insert 
locally in the center of the system flux quanta $\phi_0$ in time $T_\phi$, which 
produces an outwards Hall current $j_r \sim \Ch \ \dt \phi$ in radial direction. 
The quantization of the Hall current is related to the nontrivial Chern number $\Ch$ 
of the system. The process of inserting flux quanta increases the total angular 
momentum of the state. A more detailed discussion of the process in 
the noninteracting case can be found 
in appendix \ref{ap:FluxInsertion}. 

Starting from the $\nu = 1/2$ LN state $\ket{\LN, N}$ with $N$ particles, we insert 
two flux quanta $2 \times \phi_0$ in time $T_\phi$ in the center of the system. In 
this way, we create a two-quasi-hole excitation $\ket{2 \qh, N}$. The two-quasi-hole 
state $\ket{2\qh, N}$ already lies in the same angular momentum $L_z$ sector as the 
LN state $\ket{\LN, N+1}$ with $N+1$ particles (compare eqns. 
\eqref{eq:AngularMomentumLNstate}, \eqref{eq:AngularMomentumQuasiHoleState}).

\subsubsection{Coherent pump}

In the next step, we refill the quasi-hole excitation using a coherent pump, which 
couples a reservoir of bosons to the hole excitation. As this is done in the 
center, no angular momentum is transferred. To this end, we supplement the FQH 
Hamiltonian \eqref{eq:FQHE} by a term
\begin{equation}
 \label{eq:ContinuumProtocolCoherentPump}
 \H_\Omega = \int \d^2 \vec r \ g(t) \e^{-i \omega t} \delta^2(\vec r) \psd(\vec r) 
+ 
\hc
\end{equation}
While $g(t)$ denotes the strength of the coherent pump, the driving frequency 
$\omega = \omega_c/2$ is chosen resonant with the lowest LL (LLL). In the regime of 
large magnetic fields and low temperatures, it is sufficient to project eqn. 
(\ref{eq:ContinuumProtocolCoherentPump}) to the LLL resulting in
\begin{equation}
 \label{eq:ContinuumProtocolCoherentPumpLLL}
 \P_\LLL \H_\Omega \P_\LLL = \Omega  \e^{-i \omega t} \bd_{0,0} + \hc,
\end{equation}
where we defined the Rabi-frequency $\Omega = g/\sqrt{2\pi \lB^2}$. This corresponds 
to $\Omega_\eff$ in the toy model. 

Since the coherent pump does not change the angular momentum $L_z$ sector, we obtain 
a similar blockade mechanism as in the toy model (see Fig. 
\ref{fig:ContinuumProtocol}a with $L_z = 0,2$). In the corresponding sector, there 
is only one zero-interaction energy eigenstate, the $N+1$ LN state. The energy 
offset to any other states in the $(N+1)$-particle  manifold is of the order of 
Haldanes zeroth pseudopotential $V_0$ \cite{Haldane1983}. 
We require $\Omega \ll \omega_c,V_0$ to avoid excitations to high energy states.

\begin{figure}[t]
\centering
\epsfig{file=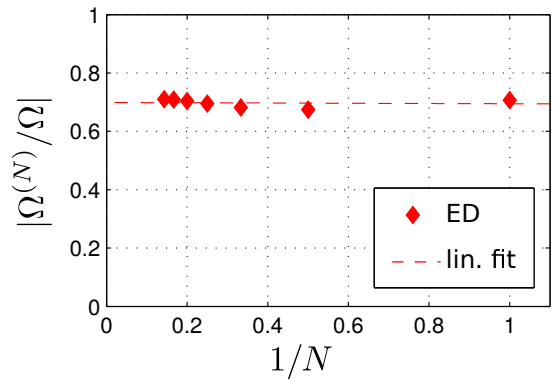, width=0.4\textwidth}
\caption{(Color online) Franck-Condon factors $\Omega^{(N)}/\Omega$, as defined in 
eqn. (\ref{eq:ContinuumFCF}), for different particle numbers $N$ calculated using 
exact diagonalization (ED) in a system with contact interaction for up to $N=7$ 
particles. A linear fit is used to extrapolate $N\rightarrow \infty$.}
\label{fig:Franck-Condon-Continuum}
\end{figure}

To insert a single particle, we use a $\pi$-pulse of duration $T_\Omega = 
\pi/2\Omega^{(N)}$. 
Here, the bare Rabi-frequency $\Omega$ is reduced by a Franck-Condon factor 
\begin{equation}
 \label{eq:ContinuumFCF}
 \Omega^{(N)}/ \Omega = \bra{\LN,N} \bd_{0,0} \ket{2\qh,N-1},
\end{equation}
which accounts for the overlap between the initial and final state. In Fig. 
\ref{fig:Franck-Condon-Continuum} we show the Franck-Condon factors for different 
particle number $N$ and extrapolate to $N \rightarrow \infty$. To this end, we 
calculated the LN wavefunction using exact diagonalization \footnote{Exact 
diagonalization is used to obtain the coefficients of the monomials 
$z_1^{m1}\ldots z_N^{m_N}$ in the Laughlin state represented in the angular momentum 
basis. Another approach would be 
to use Jack polynomials \cite{Bernevig2008}}. A linear fit suggests 
that even in the case $N \rightarrow \infty$ the Franck-Condon factor does not vanish 
with $\Omega^{(\infty)}/\Omega \simeq 0.7$.

\subsubsection{CF picture}

The first few steps of the protocol are illustrated in Fig. 
\ref{fig:ContinuumProtocol}a. Another description of the protocol, which provides a 
much simpler physical picture uses the CF picture. Firstly, due to local flux 
insertion (see Fig. 
\ref{fig:ContinuumProtocol}d), we generate a quasi-hole excitation. Secondly, the 
hole excitation is refilled by an effective coherent CF pump, for a particle 
together with one flux quantum. Analogously to the blockade mechanism, one CF 
refills the empty $\l = 0$ orbital (see Fig. \ref{fig:ContinuumProtocol}c). 
By subsequent repetition, we grow the filling $\nu^* = 1$ integer quantum Hall 
state of noninteracting CFs in a reduced magnetic field $B^*$. 

The CF picture provides an explanation, why it is possible to grow highly correlated 
LN states. Since the CF wavefunction $\Phi_\CF^{(\nu^* = 1)}$ is a separable Slater 
determinant of noninteracting CFs, we can grow the highly correlated LN state by 
adding CFs one by one to the system. Note that, although the wavefunction is a 
separable product state of CFs occupying Wannier functions (or Landau level 
orbitals), it has non-local topological order. This is because Wannier functions 
(or orbitals) of a single LL are non-local \cite{Brouder2007}. On first glance, 
this may look contradicting as the dynamical growing scheme presented in this paper 
involves only local processes. However, the non-local topological order in the system 
is generated by the Thouless pump (flux insertion).

\subsection{Performance}\label{subsec:ContinuumPerformance}

For interesting applications such as measuring the braiding statistics of elementary 
excitations, the LN 
state has to be prepared with high accuracy. To quantify the quality of the scheme, 
we calculate the fidelity $\F_N$ of being in the LN state with $N$ particles after 
$N$ steps of the protocol. The fidelity is defined as
\begin{equation}
 \label{eq:DefinitionFidelity}
 \F_N = | \bra{\Psi_N} \LN, N \rangle |,
\end{equation}
where $\ket{\Psi_N}$ is the state after $N$ steps of the protocol. In the 
calculation of the fidelity $\F_N$ we include besides nonadiabatic transitions in 
the flux insertion and coherent pump process, a particle loss rate $\gamma$. This is 
important, since decay usually plays a crucial role, in particular in photonic 
systems. 

The total time for a full cycle (flux insertion time $T_\phi$, coherent pump time 
$T_\Omega$) is $T = T_\phi + T_\Omega$. As will be discussed 
in detail later, we find for the fidelity perturbatively in the limit $\gamma T, 
(\Delta_\LN T_\phi)^{-1}, (\Delta_\LN T_\Omega)^{-1} \ll 1$
\begin{equation}
 \label{eq:Fidelity}
 \F_N \simeq \exp \L - \frac{1}{2} N \L \frac{1}{2} \gamma T (N+1) + 
\frac{\Lambda_N^2}{(\Delta_\LN T)^2} \R  \R.
\end{equation}
We see, that the losses and the nonadiabatic transitions contribute competitively 
in eqn. (\ref{eq:Fidelity}). While it is favorable to run the protocol as fast as 
possible to avoid losses, the adiabaticity requires long time scales $T$. 

For fixed 
fidelity $\F_N = 1 - \varepsilon$, $\varepsilon \ll 1$, we calculate the maximal 
number of particles $N_\max$ which can be grown with optimal period 
$T_{\mathrm{opt}}$. To leading order 
in $N$, we approximately obtain
\begin{align}
 \label{eq:FidelityNT}
 N_\max & \simeq 1.365 \varepsilon^{3/5} \L \frac{\Delta_\LN}{\Lambda_N \gamma} 
\R^{2/5}
\\
T_\mathrm{opt} & \simeq 1.431 \L \varepsilon \gamma \R^{-1/5} 
\L \frac{\Lambda_N}{\Delta_\LN} \R^{4/5}.
\end{align}
The LN state with fildelity $\F_N = 1 - \varepsilon$ can be grown in time $T_0 = 
N_\max T_{\mathrm{opt}} \sim N_\max^{3/2} \varepsilon^{-1/2} 
\Lambda_N/\Delta_\LN$ scaling only slightly faster than linear with particle number 
$N_\max$. Note, that the fidelity decreases with increasing time $t > T_0$. In the 
following, we discuss the different contributions to the fidelity (\ref{eq:Fidelity}) 
in detail. 

\subsubsection{Loss}

We assume a particle loss rate $\gamma \ll 1/T$, i.e. after one full period $T$ the 
probability of loosing a particle is small. The probability of a single decay 
process after 
$N$ steps of the protocol is then given by
\begin{equation}
 P_\gamma = 1- \exp \L - \gamma T \sum_{n=1}^N n  \R \simeq \frac{1}{2} \gamma T N 
\L N+1 \R.
\end{equation}
To leading order in $N$, the probability of loosing one particle increases 
quadratically with the particle number $N$.

\subsubsection{Flux insertion}

The process of flux insertion, as described in \ref{subsec:ContinuumProtocol}, 
introduces one flux quantum $\phi_0$ in the center of the system. For 
simplicity, we assume the flux $\phi(t) = \phi_0 \ t/T_\phi$ to change linear in 
time $t$. In the fully adiabatic protocol, the angular momentum $\l$ is therefore 
increased by one without coupling different LLs $n,n'$. We calculate in the 
noninteracting case the nonadiabatic coupling matrix element $\kappa = \langle n',\l 
| -i \partial_\l | n,\l \rangle$ between different LLs. To estimate the 
scaling of the probability 
$P_\phi$ for excitation of higher LLs, we consider the coupling between the LLs $n$ 
and 
$n+1$. In the regime of long times $T_\phi$, we determine $P_\phi$ perturbatively 
(see appendix \ref{ap:FluxInsertionFidelity}). Using the Laughlin gap $\Delta_\LN$, 
we obtain
\begin{equation}
 \label{eq:FluxInsertionProbability}
 P_\phi \simeq \frac{\tilde \kappa^2}{(\Delta_\LN T_\phi)^2},
\end{equation}
where $\tilde \kappa$ is a nonuniversal coupling constant. Importantly, nonadiabatic 
transitions to higher LLs scale as $\sim 1/(\Delta_\LN T_\phi)^2$.

\subsubsection{Coherent pump}

\begin{figure}[t]
\centering
\epsfig{file=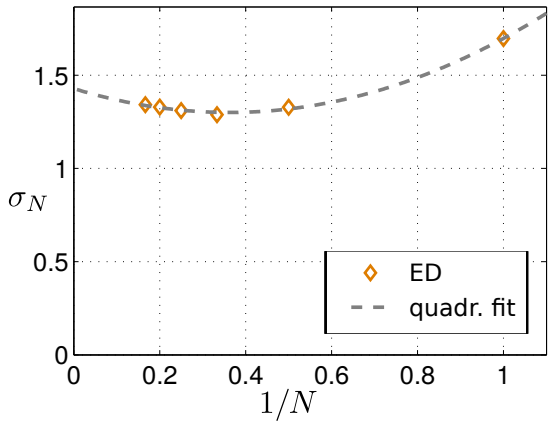, width=0.4\textwidth}
\caption{(Color online) Scaling of the nonuniversal factor $\sigma_N$ from eqn. 
(\ref{eq:CoherentPumpProbability}) with particle number $N$. We interpolate $N 
\rightarrow \infty$ with a quadratic fit. } 
\label{fig:FidelityCoherentPump}
\end{figure}

The coherent pump (\ref{eq:ContinuumProtocolCoherentPumpLLL}), as discussed in 
\ref{subsec:ContinuumProtocol}, couples in zeroth order in $\Omega/\Delta_\LN$ only 
the two-quasi-hole state $\ket{2\qh,N}$ to the Laughlin state $\ket{\LN,N+1}$. In 
first order perturbation theory, couplings to excited states in the $N-1, N, N+1, 
N+2$ 
particle sectors are relevant. We obtain for the probability of nonadiabatic 
excitations 
\begin{equation}
 \label{eq:CoherentPumpProbability}
 P_\Omega \simeq \frac{\sigma_N^2}{(\Delta_\LN T_\Omega)^2},
\end{equation}
where we used the $\pi$-pulse time $T_\Omega = \pi/2\Omega$. The nonuniversal factor 
$\sigma_N$ includes Franck-Condon factors and excitation energies $\Delta^{(N)}$ 
of undesired states weighted by the Laughlin gap $\Delta_\LN$ (see appendix 
\ref{ap:CoherentPumpFidelity}). In Fig. \ref{fig:FidelityCoherentPump} $\sigma_N$ is 
calculated for various particle numbers $N$ using exact diagonalization. For 
$N\rightarrow 
\infty$ we extrapolate with a quadratic fit $\sigma_\infty \simeq 1.4$. 

To conclude, we obtain the estimate for the fidelity in eqn. (\ref{eq:Fidelity}), 
summarizing the contributions of flux insertion $P_\phi$ and coherent pump 
$P_\Omega$ for fixed ratio $T_\Omega/T_\phi$. Assuming, the coefficients $\tilde 
\kappa, \sigma_N$ to depend only slightly on the particle number $N$, we define the 
factor $\Lambda_N$ containing all constant contributions. Then, using
\begin{equation}
 \mathcal{F}_N = \exp \L -\frac{1}{2} \L P_\gamma + P_\phi + P_\Omega \R \R
\end{equation}
we obtain eqn. (\ref{eq:Fidelity}).


\section{Lattice and Fractional Chern Insulators}\label{sec:Lattice}

Lattice systems are promising candidates to realize FQH physics, 
since the magnetic fields realized in these systems are very strong, such that low 
magnetic filling factors $\nu \leq 1$ can be achieved while keeping a sufficiently 
large density of particles. However, in this case the magnetic length $\lB$ and the 
lattice constant $a$ become of comparable size. Therefore lattice effects, such as 
dispersive bands, play a crucial role. Now, we introduce an effective CF 
lattice model, which allows us to include these effects in our investigation. 
Moreover, we consider finite systems now, where edge states are present. 

\subsection{Hofstadter Hubbard Model}
We consider a two dimensional lattice described in a tight-binding model with 
nearest neighbor hopping terms $J$. The Peierls phases are chosen to mimic 
a magnetic field in Landau gauge. Additionally, we consider onsite 
interactions $U$ resulting in the Hofstadter Hubbard Hamiltonian
\begin{multline}
\label{eq:HofstadterHubbard}
 \H = - J \sum_{x,y} \left[ \ad_{x+1,y}\a_{x,y} \e^{-i 2\pi \alpha y} + 
\ad_{x,y+1}\a_{x,y} + \hc \right] \\
+ U/2 \sum_{x,y} \ad_{x,y}\ad_{x,y}\a_{x,y}\a_{x,y}.
\end{multline}
The $x,y$ coordinates are measured in units of the lattice constant 
$a$. We use the operators $\a_{x,y}$ to denote the bosonic annihilation of a 
particle 
at site ($x,y$). The flux per plaquette $\alpha$ (in units of the flux quantum) 
accounts for the magnetic field 
penetrating the two dimensional lattice. In particular, it sets the magnetic length 
$\lB = a/\sqrt{2\pi \alpha}$, which describes the extend of the cyclotron orbits. 

In recent experiments, the realization of the Hofstadter model was shown in a 
photonic system \cite{Hafezi2013} as well as ultracold gases \cite{Aidelsburger2013, 
Miyake2013}. The photonic system \cite{Hafezi2013} implements a tight-binding model 
using ringresonators in the optical wavelength regime. The tunneling is achieved 
using off-resonant ringresonators, which have a different length for hopping forward 
and backward and the Peierls phases are determined by the optical path difference. 
In experiments 
with ultracold gases \cite{Aidelsburger2013, Miyake2013}, standing waves are used to 
create a two dimensional optical lattice. Due to a strong field gradient along one 
direction, the particles are localized. Using laser-assisted tunneling techniques, 
the hopping elements are implemented with an additional Peierls phase. In the 
experiments, the Peierls phase can be engineered to realize e.g. a flux per 
plaquette 
of $\alpha = 1/4$.

Note, that the continuum limit $\alpha \rightarrow 0$, where the magnetic length 
$\lB$ is much larger than the lattice constant $a$, corresponds to the FQH model 
\eqref{eq:FQHE}.

\subsection{Ground State and Excitations}\label{subsec:LatticeGroundState}

We summarize the properties of the ground state of the Hofstadter Hubbard model 
(\ref{eq:HofstadterHubbard}) for bosons with magnetic filling factor $\nu = 1/2$ in 
two different geometries. In the case of a \textit{torus} discussed in 
\cite{Sorensen2005, Hafezi2007}, the exact ground state was compared to the Laughlin 
state (\ref{eq:LNwavefunction}) projected on a lattice for different flux per 
plaquette $\alpha$. It was found, that the Laughlin wavefunction provides a good 
description up to $\alpha \simeq 0.2$. Moreover, the many-body Chern number remains 
$\Ch = 1/2$ until the flux per plaquette reaches a critical value of $\alpha \simeq 
0.4$. The size of the Laughlin gap $\Delta_\LN$ depends on the parameters $\alpha$ 
and $U$. For instance, at $\alpha \simeq 0.1$, the Laughlin gap saturates at 
$\Delta_\LN \simeq 0.1J$ for large $U \gg J$. 

\begin{figure}[t]
\centering
\epsfig{file=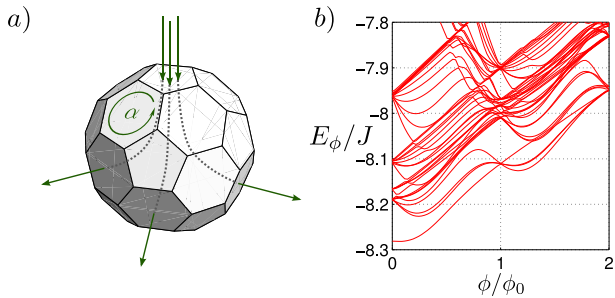, width=0.45\textwidth}
\caption{(Color online) a) Flux insertion on a $\mathrm{C}_{60}$ buckyball geometry. 
b) Many-body spectrum $E(\phi)$ on a $\mathrm{C}_{60}$ buckyball starting with $N=3$ 
particles and $N_\phi=4$ flux quanta and introducing two more flux quanta $2 
\times \phi_0$. 
} 
\label{fig:LatticeC60GroundState}
\end{figure}

In \cite{Grusdt2014a}, we numerically analyzed the ground state in a spherical 
geometry using a buckyball lattice with $N_s = 60$ sites (see Fig. 
\ref{fig:LatticeC60GroundState}a). To realize a filling $\nu = 1/2$ in the continuum 
on a sphere, the relation between particle number $N$ and flux quanta $N_\phi$ is 
$N_\phi = 2(N-1)$. We identify the topological order of the ground state by 
inserting two flux quanta $\phi_0$ in a system with $N=3$ particles and $N_\phi = 4$ 
flux quanta. The many-body spectrum $E_\phi$ during flux insertion is shown in Fig. 
\ref{fig:LatticeC60GroundState}b. We observe a single incompressible ground 
state with many-body gap $\Delta \simeq 0.1J$ for $U = 10J$, similar to the case on 
a torus. Importantly, we obtain the correct counting of nearly degenerate 
quasi-holes states after inserting one and two flux quanta as expected from the 
continuum limit on a sphere. In both lattice cases we expect the ground state to be 
in the same topological universality class as the LN state in the continuum.

\subsection{Effective CF Lattice Model}\label{subsec:LatticeCFModel}
Before we introduce the effective CF lattice model, we briefly summarize the results 
of the exact simulations in reference \cite{Grusdt2014a}. We implemented the 
protocol on a $\mathrm{C}_{60}$ lattice up to $N=3$ particles. Additionally, we used 
a similar quasi-hole trapping potential as in eqn. (\ref{eq:ToyModelHoleTrapping}). 
The numerical results show, that the state $\ket{\Psi(t)}$ prepared after three steps 
of the protocol is close to the LN ground state $\ket{\mathrm{gs}}$ with $N=3$ 
particles. Moreover, as expected from the blockade mechanism in the coherent pump, 
the number fluctuations $\Delta N(t)$ are small throughout the protocol. 

To discuss edge effects, larger systems with many particles are required, such that 
bulk and edge states can be distinguished. However, using exact diagonalization only 
a few particles are feasible on large lattice systems with $N_\mathrm{s} \gtrsim 60 
$ 
sites. To simulate such a system, we need an effective model describing the low 
energy dynamics of the Hamiltonian (\ref{eq:HofstadterHubbard}). 

In the spirit of the 
composite fermion principle, we assume a model of noninteracting composite fermions 
on a lattice. By attaching one flux quantum to a boson, as shown in Fig. 
\ref{fig:ContinuumProtocol}b, an (approximately) noninteracting composite fermion is 
formed in a 
reduced magnetic field. The simplest tight-binding model with a reduced magnetic 
field, $\alpha^* = \alpha/2$, only considers nearest neighbor hopping elements 
$J^*$. 
Under these assumptions, the effective model is
\begin{equation}
\label{eq:CFModel}
 \H_\CF = -J^* \sum_{x,y} \left[ \cd_{x+1,y}\c_{x,y} \e^{-i 2\pi \alpha^* y} + 
\cd_{x,y+1}\c_{x,y} + \hc \right],
\end{equation}
where $\cd_{x,y}$ creates a composite fermion at site ($x,y$). The only free 
parameter $J^*$ in this model determines the time scale of the dynamics. By 
comparing the bandwidth of the full many-body spectrum of interacting bosons to the 
free CF single-particle spectrum on a  $\mathrm{C}_{60}$ buckyball at $N_\phi = 6$ 
flux quanta, we find $J^* \simeq J$. 

Now, we conjecture that the effective CF lattice model describes correctly the low 
energy dynamics. Yet, there is no proof that the CF theory is applicable in a 
lattice. However, there are several hints that the essential physics can still be 
understood in terms of CFs. First of all, the CF picture in a lattice captures the 
correct counting of quasi-hole excitations \footnote{Note, that the counting of 
quasi-hole excitations on a $\mathrm{C}_{60}$ buckyball lattice and the counting 
in the continuum on the sphere are equal and thus also the CF counting is correct.}. 
Thus, the low energy excitations should be described correctly. Moreover, the LN 
wavefunction is a special case of the more general CF theory, when the lowest CF-LL 
$n^*=0$ is filled. As shown in 
\cite{Sorensen2005} the LN wavefunction projected on a lattice describes the 
many-body ground state on a lattice accurately up to relatively high flux per 
plaquette $\alpha 
\simeq 0.2$. Therefore, we limit the flux per plaquette in our effective model to 
$\alpha^* \leq 0.1$. Finally, CF states from the Jain sequence \cite{Jain1989}, 
other than the LN state, have been identified in a lattice \cite{Moeller2009,Liu2013} 
for small flux per plaquette. Therefore, we expect that this effective model 
captures the essential physics of the original many-body model in the lowest Chern 
band ($\LChB$), including the dynamics of the hole excitations. As this model is 
supposed to describe only the low energy regime, excitations to higher Chern bands 
($\HChB$s) are not expected to be described correctly.

\subsection{Numerical Results}\label{subsec:LatticeNumericalResults}


\subsubsection{Numerical implementation of the protocol}
The protocol is implemented in both, the spherical geometry on a 
$\mathrm{C}_{60}$ buckyball as in \cite{Grusdt2014a} as well as the square 
lattice with open boundary conditions. We use the method explained in 
\cite{Grusdt2014a} to insert flux quanta locally. 

Unlike the coherent boson pump \eqref{eq:ContinuumProtocolCoherentPump}, we model 
the coherent CF pump by coupling a CF reservoir mode to the central site of the 
system. The reservoir mode is refilled in each step of the protocol. Therefore, we 
insert at most one CF per cycle. This is similar to the blockade mechanism for 
bosons and therefore only justified in the limit of large interactions $U \gg J$, 
where corrections scale as $(\Omega/\Delta_\LN)^2$ (see eqn. 
\eqref{eq:CoherentPumpProbability}).

Moreover, we implement the quasi-hole trapping potential by including an onsite 
potential $g_\mathrm{h}$ on the central site. 

As explained for the toy model, we introduce loss channels at the boundaries of the 
system to prevent high energy excitations. Then, we calculate the time evolution of 
the correlation matrix elements $\av{\cd_{x,y}\c_{x',y'}}$. by solving the 
corresponding master equation in Lindblad form ($\hbar = 1$)
\begin{equation}
 \dt \hat \rho = -i [\H_\CF,\hat{\rho}] + \frac{1}{2} \sum_{x,y} '
 2 \hat l_{x,y} \hat \rho \hat l^\dagger_{x,y} 
 - \{ \hat l^\dagger_{x,y} l_{x,y}, \hat \rho \}
\end{equation}
numerically. Absorbing boundary conditions are described by the jump operators 
$l_{x,y} = \sqrt{\gamma_\Edge}\c_{x,y}$ with loss rate $\gamma_\Edge$ and we 
restrict the sum $\sum '$ to the edge of the system. 

Note, that the loss of a CF is a loss of both, a flux quantum and the boson it was 
attached to. Therefore, we only allow this loss term at the edge of the system, where 
the meaning of a free flux quantum without a particle is obsolete.

\subsubsection{Performance}
To investigate the performance of the protocol, we use the spherical geometry 
without edges. The bandwidth $\Delta E/J$ of the $\mathrm{C}_{60}$ buckyball with up 
to $N_\phi = 6$ flux quanta is small and thus we can neglect the effect of dispersive 
bands. In section \ref{subsec:ContinuumPerformance}, we have seen that the 
topological pump as 
well as the coherent pump in the continuum need sufficiently 
long times $T_\phi, T_\Omega \gg \Delta_\LN^{-1}$ to avoid nonadiabatic excitations. 
The excitation probability to the excitonic states scale as $P_\phi \sim 
T_\phi^{-2}, 
P_\Omega \sim T_\Omega^{-2} 
\sim \Omega^2$ (see eqns. \eqref{eq:FluxInsertionProbability}, 
\eqref{eq:CoherentPumpProbability}). Here, we show that the scaling for $P_\phi, 
P_\Omega$ also holds for the lattice in the perturbative regime.

\begin{figure}[t]
\centering
\epsfig{file=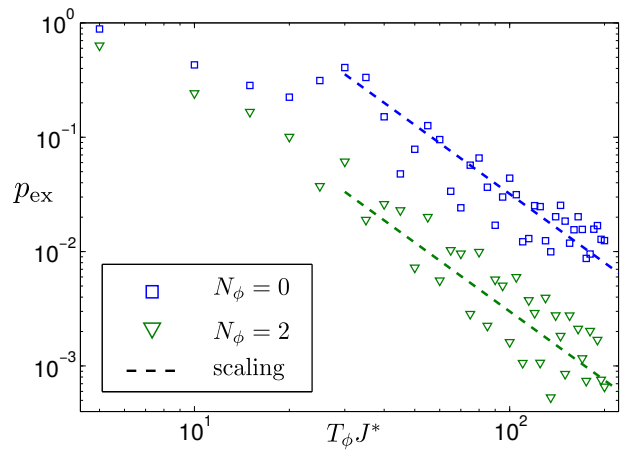, width=0.45\textwidth}
\caption{(Color online) Shown is the excitation probability $p_\mathrm{ex}$ to high 
energy states in the case of flux insertion. Starting with particle number $N=1$ 
($N=3$) in the ground 
state $\ket{\mathrm{gs}}$ at $N_\phi = 0$ ($N_\phi = 2$) flux quanta, we insert one 
flux quantum and analyze the probability $p_\mathrm{ex}$ of being in an excited 
state 
for various flux insertion times $T_\phi$. The dashed lines indicate the expected 
scaling from eqn. \eqref{eq:FluxInsertionProbability} 
in section \ref{subsec:ContinuumPerformance}. }
\label{fig:C60PerformanceFluxInsertion}
\end{figure}

Firstly, we analyze the excitation probability $p_\mathrm{ex}$ in the case of 
\textit{flux insertion}. We start from the ground state $\ket{\mathrm{gs}}$ with 
$N=1$ ($N=3$) particle(s) at $N_\phi = 0$ ($N_\phi = 2$) flux quanta and insert 
one flux quantum in time $T_\phi$ to create a CF quasi-hole excitation $\ket{\qh}$. 
Figure \ref{fig:C60PerformanceFluxInsertion} shows the probability $p_\mathrm{ex}$ 
for excitation of higher bands. Besides an oscillatory behavior with increasing 
duration $T_\phi$, we confirm the expected scaling $P_\phi \sim T_\phi^{-2}$in the 
perturbative regime. 

\begin{figure}[t]
\centering
\epsfig{file=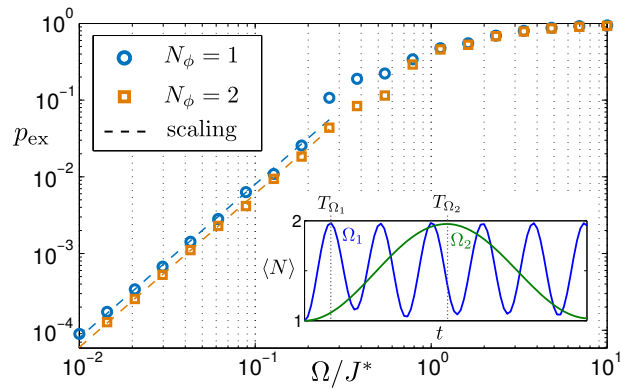, width=0.45\textwidth}
\caption{(Color online) We show the excitation probability $p_\mathrm{ex}$ to high 
energy states for different Rabi-frequencies $\Omega$ in the coherent CF pump. 
Starting 
from the quasi-hole state with $N_\phi=1$ 
($N_\phi=2$) flux quantum, we coherently couple to the ground state with $N=2$ 
($N=3$) particles for various 
Rabi-frequencies $\Omega$. The dashed lines indicate the expected scaling 
from section \ref{subsec:ContinuumPerformance}. In the inset, we extract the 
$\pi$-pulse time $T_\Omega$ when the maximal number of particles is in the system, 
for two different values of $\Omega$.  } 
\label{fig:C60PerformanceCoherentPump}
\end{figure}

To analyze the excitation probability $p_\mathrm{ex}$ for the \textit{coherent CF
pump}, we start from the quasi-hole state $\ket{\qh}$ at $N_\phi = 1$ ($N_\phi 
= 2$) flux quantum. The coherent pump is coupled resonantly to the hole excitation 
for different bare Rabi-frequencies $\Omega$. In Fig. 
\ref{fig:C60PerformanceCoherentPump} we plot $p_\mathrm{ex}$ for different $\Omega$. 
The time $T_\Omega$ needed for a $\pi$-pulse 
is determined by the maximal achievable particle number $\av{N}$ (see inset Fig. 
\ref{fig:C60PerformanceCoherentPump}). For time $T_\Omega$, we extract the 
excitation probability $p_\mathrm{ex}$ of being not in the ground state with $N=2$ 
($N=3$) particles. We find excellent agreement with the expected 
scaling in the perturbative regime. 

\subsubsection{Dispersive bands -- quasi-hole trapping}
As noted in the toy model section \ref{sec:GrowingScheme}, the finite bandwidth 
leads to an intrinsic dispersion of the quasi-hole excitations. Without trapping the 
hole excitations, the coherent pump cannot replenish them efficiently. 

\begin{figure}[t]
\centering
\epsfig{file=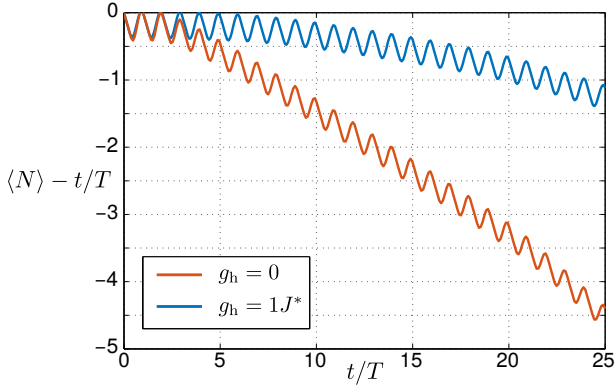, width=0.45\textwidth}
\caption{(Color online) Shown is the number of particles $\av{N}$ during the 
protocol in time $t/T$ in a system with and without trapping potential 
$g_\mathrm{h}$. The 
lattice size is $11 \times 11$ with an effective flux per plaquette $\alpha^* = 
0.1$. We insert flux in time $T_\phi = 60/J^*$. The $\pi$-pulse time $T_\Omega$ for 
$g_\mathrm{h}=0$ ($g_\mathrm{h}=1J^*$) is $T_\Omega = 100/J^*$ ($T_\Omega = 
130/J^*$). } 
\label{fig:PlanarDispersiveBands}
\end{figure}

We numerically analyze the effect of dispersive bands on a square lattice. The 
results are shown in Fig. \ref{fig:PlanarDispersiveBands}, where we compare the 
number of CFs $\av{N}(t)$ for 25 cycles of the protocol with and without quasi-hole 
trapping potential $g_\mathrm{h}$. Here, $T = T_\phi + T_\Omega$ is the time needed 
for one step of the protocol. The trapping potential improves the efficiency of the 
growing scheme already after three steps. Therefore, we include the 
quasi-hole trapping potential $g_\mathrm{h}$ for the following discussions.

\subsubsection{Finite systems -- edge decay}
Let us finally discuss the effects of a finite system, where edge states are 
present. 
As 
explained in the toy model section \ref{sec:GrowingScheme}, during the protocol edge 
states will transport particles to high energy states. To reach a homogeneous 
particle density in the bulk of the $\LChB$, absorbing boundaries 
can be implemented to prevent excitations to $\HChB$s. 

\begin{figure}[t]
\centering
\epsfig{file=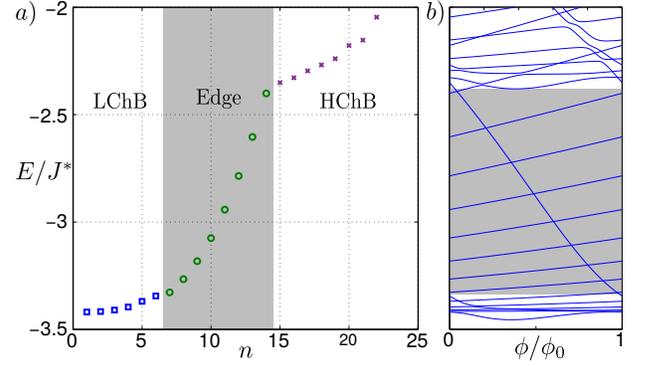, width=0.45\textwidth}
\caption{(Color online) a) Low energy spectrum of the CF Hamiltonian 
(\ref{eq:CFModel}) with $\alpha^* = 0.1$ and lattice size $11 \times 11$. States are 
labeled with number $n$. We identify three different sectors using additionally the 
density distribution of the states: Lowest Chern band ($\LChB$) $n = 1, \ldots, 6$, 
edge states ($\Edge$) $n=7,\ldots, 14$ and higher Chern bands ($\HChB$s) $n\geq 15$. 
b) Energy spectrum during insertion of 
one flux quantum $\phi_0$ with parameters as in a).  } 
\label{fig:LatticeEnergyFluxInsertion}
\end{figure}

In the case of open boundary conditions, we identify three different regimes in the 
single particle spectrum of the CFs, shown in Fig. 
\ref{fig:LatticeEnergyFluxInsertion}a. Due to the finite size of the system, edge 
states occur between the states of the $\LChB$ and those of the $\HChB$. Crucially, 
the effective CF model on the lattice is not supposed to describe the dynamics of 
the 
high energy states correctly. To prepare a LN type state in the bulk of a finite 
systems, it is necessary to avoid high energy excitations.

The free CF energy spectrum of a finite system with trapping potential 
$g_\mathrm{h}$ 
during adiabatic flux insertion is depicted in Fig. 
\ref{fig:LatticeEnergyFluxInsertion}b. Besides the creation of a hole excitation in 
the $\LChB$, edge states occur connecting the low and high energy states. Due to 
the edge states, particles will be excited to the $\HChB$s of the system during the 
protocol. 

\begin{figure}[t]
\centering
\epsfig{file=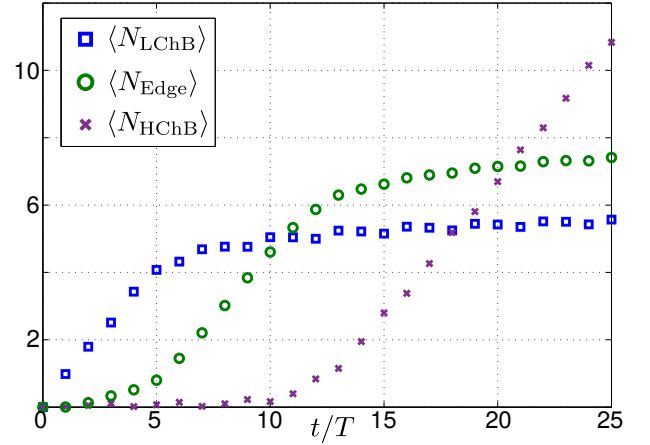, width=0.45\textwidth}
\caption{(Color online) Particle number $\av{N_\LChB}$, $\av{N_\Edge}$, 
$\av{N_\HChB}$ in the three different energy sectors, defined in Fig. 
\ref{fig:LatticeEnergyFluxInsertion}a, after each cycle in the 
protocol. The parameters used here are the same as in Fig. 
\ref{fig:PlanarDispersiveBands} with trapping potential $g_\mathrm{h}$. } 
\label{fig:PlanarParticleNumber}
\end{figure}

\begin{figure}[t]
\centering
\epsfig{file=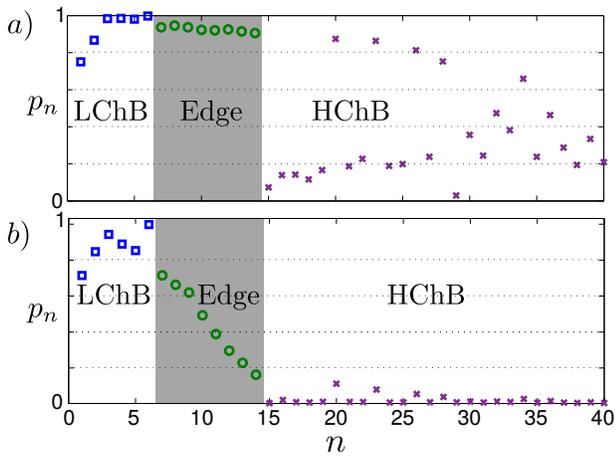, width=0.45\textwidth}
\caption{(Color online) Population $p_n$ of the low energy CF states at the end of 
the protocol. a) Parameters as in Fig. \ref{fig:PlanarDispersiveBands} after 25 
steps of the protocol and b) after 35 steps of the protocol with absorbing boundary 
conditions ($\gamma_\Edge = 0.0025 J^*$).  } 
\label{fig:PlanarEndProtocol}
\end{figure}

This is shown in Fig. \ref{fig:PlanarParticleNumber}, where we analyze the particle 
number $\av{N_\LChB}, \av{N_\Edge}, \av{N_\HChB}$ in the three different regions, 
defined in Fig. \ref{fig:LatticeEnergyFluxInsertion}a, after each step of the 
protocol. As expected, in the first few steps the number of 
particles in the $\LChB$ $\av{N_\LChB}$ increases. However, before the $\LChB$ is 
completely filled, edge states are populated and as a consequence the number of 
states in $\HChB$s $\av{N_\HChB}$ starts to increase. Figure 
\ref{fig:PlanarEndProtocol}a shows the population $p_n$ of the first CF states  
(labeled with integer $n$) after 25 steps of the protocol.

To avoid the population of $\HChB$s, we include absorbing boundaries as discussed 
in the toy model. Since edge 
states are localized at the boundaries of the system, the bulk properties of the 
system will only slightly be affected. However, for a properly chosen decay rate 
$\gamma_\Edge$ on the boundary, edge excitations will be lost before they reach the 
$\HChB$ 
during the protocol. In 
Fig. \ref{fig:PlanarEndProtocol}b we show the population of the first CF states 
after 35 steps of the protocol. The $\HChB$s are only very weakly populated due to 
the 
absorbing boundary conditions. The CF density $\rho^*$ of the last cycle is shown 
in Fig. \ref{fig:LatticeDensity_Protocol}. In terms of the CFs we reach an average 
CF filling factor of $\nu^* \simeq 0.9$ of the $\LChB$, which is close to the 
optimal 
value $\nu^* = 1$.

\section{Summary \& Outlook}

In conclusion, we discussed a protocol which allows to grow topologically ordered 
states in interacting many-body systems. We explained all necessary 
ingredients using the SLBHM as a simple toy model. We showed that in the flat-band 
limit a combination of a topologically protected Thouless pump \cite{Thouless1982}, 
creating a local hole excitation, and a coherent pump, refilling the hole 
excitation, are sufficient. Moreover, we extended the protocol to the case of 
dispersive bands and finite systems with edges.

Furthermore, we discussed the protocol in detail in both, the continuum case and the 
lattice case of fractional quantum Hall systems. In the continuum, we estimated the 
fidelity of the protocol depending on the particle number $N$. To describe 
numerically 
large lattice systems with many particles, we introduced an effective model based on 
the CF description of the FQHE. This allows us to simulate large systems and include 
edge effects. We showed, that a quasi-hole trapping potential can be 
used in the case of dispersive bands to maintain a high efficiency of the protocol. 
Moreover, absorbing boundaries are used in finite systems, where edge states would
transport excitations to higher bands. We showed that in the case of dispersive 
bands 
and even in the presence of edge states, a high CF filling factor $\nu^* \simeq 0.9$ 
is achievable, in large systems with more than 10 particles.

We believe, that our protocol can be used to grow other exotic states than the LN 
state, e.g. the Moore-Read Pfaffian \cite{Moore1991}. So far, we did not consider 
experimental realizations. However, ultracold gases as well as photonic systems are 
promising candidates. Moreover, the efficiency of our scheme can be increased by 
introducing multiple pairs of topological and coherent pumps.

\section*{Acknowledgements}
The authors would like to thank M. Hafezi and L. Glazman for stimulating 
discussions. 
F.G. is a recipient of a fellowship through the Excellence Initiative (DFG/GSC 266) 
and is grateful for financial support from the "Marion K\"oser Stiftung". 

\appendix

\section{Flux Insertion in the IQHE}\label{ap:FluxInsertion}
To address the problem of flux insertion, we briefly review the Landau Level (LL) 
problem. To this end, we use a basis, which is not typically used in standard 
textbooks. It will turn out, that this basis allows a simple understanding of the 
flux insertion process.

\subsection{Landau Level}
\begin{figure}[t]
\centering
\epsfig{file=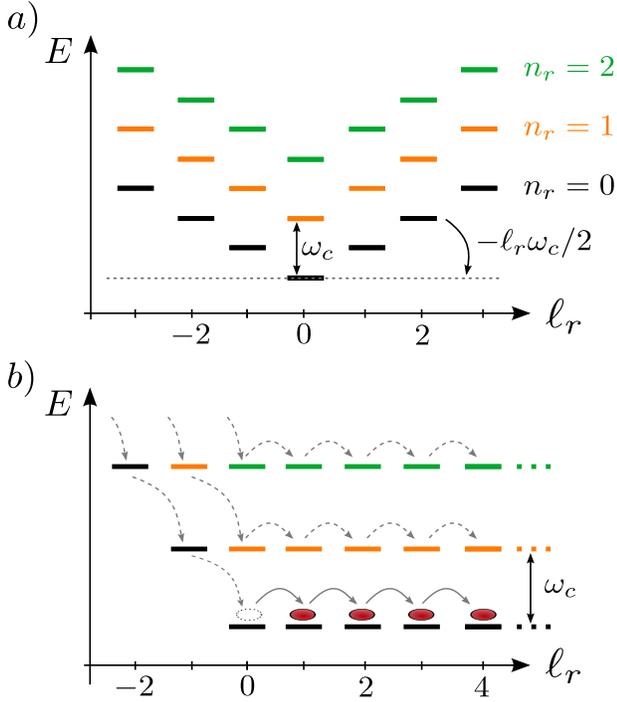, width=0.45\textwidth}
\caption{(Color online) a) Schematic picture of the Landau Level construction using 
the two dimensional harmonic oscillator basis $\ket{n_r,\l_r}$. b) In the flux 
insertion process, the angular momentum $\l_r$ is increased by one, while keeping 
the quantum number $n_r$ fixed. } 
\label{fig:ap_FluxInsertion}
\end{figure}

The LL Hamiltonian in symmetric gauge, $\vec A = B/2 (-y,x,0)$, is ($\hbar = 1$)
\begin{align}
\label{eq:ap_LL}
 \H_0 &= \frac{1}{2M} \L \vec p - \vec A \R^2 \\
&= \frac{1}{2M} \vec p^2 + \frac{1}{2} M \L \frac{\omega_c}{2} \R^2 \L x^2 + y^2 \R 
- \frac{\omega_c}{2} L_z. \nonumber
\end{align}
As angular momentum $L_z$ is a conserved quantity, i.e. $[\H_0, L_z] = 0$ , we can 
use the eigenbasis $\ket{n_r, \l_r}$ of a two dimensional harmonic oscillator. Here, 
$n_r = 0,1,\ldots$ corresponds to the energy levels of the harmonic oscillator and 
$\l_r \in \mathbb{Z}$ to those of the angular momentum. We obtain 
\begin{align}
 L_z \ket{n_r, \l_r} &= \l_r \ket{n_r, \l_r} \\
\H_0 \ket{n_r, \l_r} &= \frac{\omega_c}{2} \L 2 n_r + |\l_r| - \l_r \R \ket{n_r, 
\l_r}.
\end{align}
The energy spectrum of both, the harmonic oscillator and the LL, are shown in Fig. 
\ref{fig:ap_FluxInsertion}.

\subsection{Flux insertion}
In 1981 Laughlin \cite{Laughlin1981} explained the quantization of the Hall current 
using the argument of flux insertion. This idea can be used to create localized hole 
excitations in the quantum Hall effect. The idea is as follows. After introducing 
one flux quantum $\phi_0$ adiabatically in the center, a circular electric field is 
generated. The Hall response thus leads to a radial 
outwards current creating a hole in the center of the system. As shown below, the 
quasi-hole excitation is quantized.

To realize Laughlin's argument, we include in eqn. 
\eqref{eq:ap_LL} the vectorpotential
\begin{equation}
 \vec A_\phi = - \frac{\phi (t)}{2\pi r} \vec e_\varphi.
\end{equation}
Defining a new angular momentum
\begin{equation}
 L_z^{'} = L_z + \phi/\phi_0,
\end{equation}
we obtain the same structure as in eqn. \eqref{eq:ap_LL}. However, by adiabatically 
increasing $\phi(t)$, we change the angular momentum of the system. By inserting 
one flux quantum $\phi_0$, we increase the angular momentum $\l_r$ of all states by 
one, while the quantum number $n_r$ stays fixed. This generates the spectral flow 
depicted in FIG. \ref{fig:ap_FluxInsertion}b.

\section{Nonadiabatic Corrections in the Flux Insertion 
Process}\label{ap:FluxInsertionFidelity}
To calculate the probability $P_\phi$ of exciting particles to higher LLs during 
flux insertion in the noninteracting case, we use the basis discussed in appendix
\ref{ap:FluxInsertion}. By inserting adiabatically one flux quantum $\phi(t) =  
\phi_0 \ t/T_\phi$ in time $T_\phi \gg 1 / \omega_c$, the angular momentum 
$\l_r(t)$ increases by one, i.e. $\l_r(T_\phi) = \l_r(0) + 1$. Therefore, starting 
from state $\ket{n_r,\l_r(0)}$, we end in the state $\ket{n_r,\l_r(0)+1}$. Here, we 
calculate perturbatively in the regime $(\omega_c T_\phi)^{-1} \ll 1$ the scaling of 
the probability $P_\phi$ of exciting particles to higher LLs. 

\begin{figure}[t]
\centering
\epsfig{file=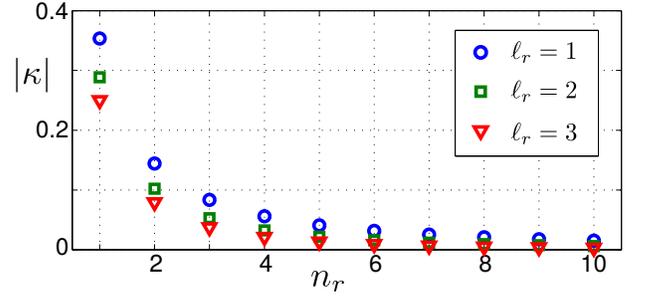, width=0.45\textwidth}
\caption{(Color online) Coupling $|\kappa|$ from the lowest LL $n_r' = 0$ to higher 
LLs $n_r$ for different angular momentum $\l_r$.  } 
\label{fig:ap_Coupling_kappa}
\end{figure}

The nonadiabatic coupling $g_\phi$ between different LLs $n_r \neq n_r'$ is
\begin{align}
 g_\phi &=  \langle n_r', \l_r | -i \dt |n_r, \l_r \rangle \nonumber \\
&= \langle n_r', \l_r | -i \partial_{\l_r} |n_r, \l_r \rangle /T_\phi = \kappa 
/T_\phi.
\end{align}
In Fig. \ref{fig:ap_Coupling_kappa}, we plot the coupling $|\kappa|$ from the lowest 
LL $n_r' = 0$ to higher LLs $n_r$ for different angular momentum $\l_r$. As 
expected, the coupling to higher LLs decreases.

To estimate the scaling of $P_\phi$ with flux insertion time $T_\phi$ in the 
perturbative regime $(\omega_c T_\phi)^{-1} \ll 1$, we consider a simple two level 
approximation with LLs $n_r, n_r+1$  and constant coupling $\kappa$. Starting from 
the state $\ket{n_r,\l_r}$, we calculate in first order perturbation theory the 
probability for ending in state $\ket{n_r+1,\l_r}$. We obtain approximately
\begin{equation}
 P_\phi \simeq \frac{\kappa^2}{(\omega_c T_\phi)^2} 2 \L 1-\cos(\omega_c T_\phi) \R.
\end{equation}
We expect, that the scaling of $P_\phi$ with $T_\phi$ in the interacting case to be 
the same as in the noninteracting case, when the cyclotron frequency $\omega_c$ is 
replaced by the many-body gap $\Delta_\LN$. This leads to eqn. 
\eqref{eq:FluxInsertionProbability}.

\section{Nonadiabatic Corrections in the Coherent Pump 
Process}\label{ap:CoherentPumpFidelity}

\begin{figure*}[t]
\centering
\epsfig{file=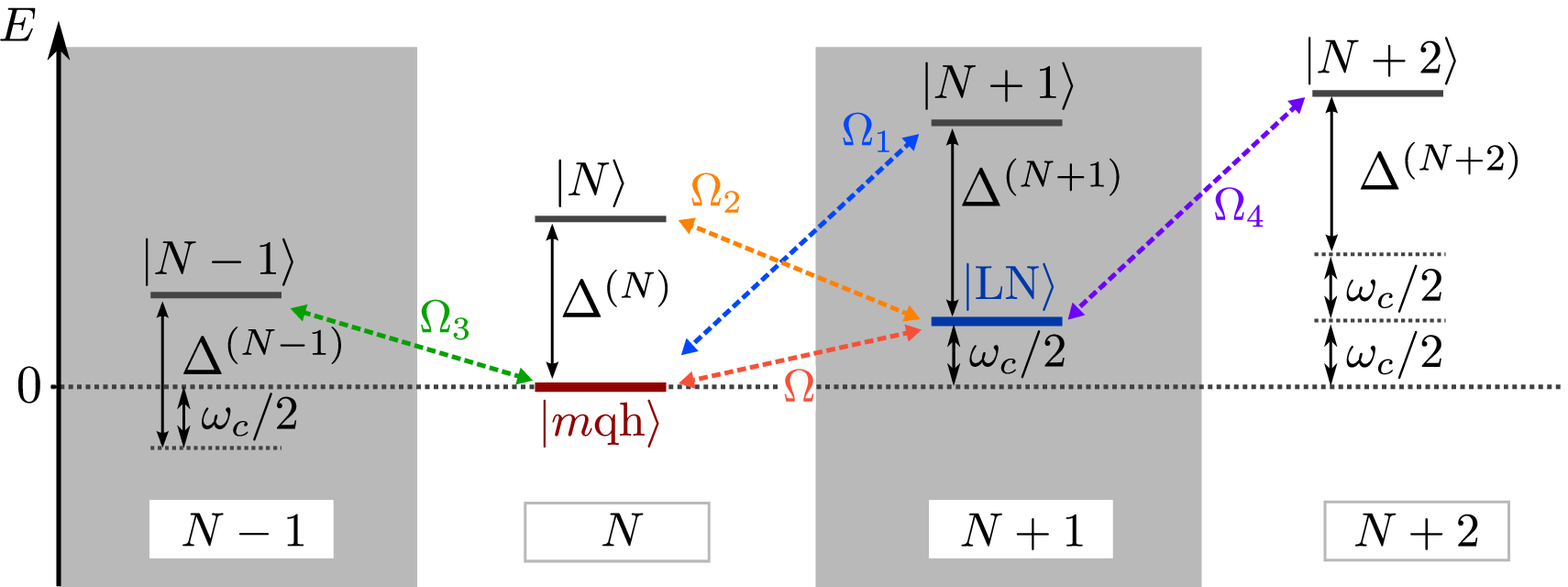, width=0.7\textwidth}
\caption{(Color online) Schematic overview over the coupling between states 
in different particle sectors in first order in $\Omega/\Delta_\LN$. 
Starting from the quasi-hole state $\ket{\qh}$ of $N$ particles or the LN state 
$\ket{\LN}$ with $N+1$ particles, the particle sectors from $N-1$ to $N+2$ are 
involved. $\Omega_i$ denote the many-body Rabi-frequencies reduced by their 
Franck-Condon factors with respect to the bare Rabi-frequency $\Omega$. The 
many-body 
gaps $\Delta^{(N)} \geq \Delta_\LN$ to the excitonic states $\ket{N}$ are multiples 
of the LN gap $\Delta_\LN$. } 
\label{fig:ap_CoherentPump}
\end{figure*}
In zeroth order in $\Omega/\Delta_\LN \ll 1$, the coherent pump couples the 
quasi-hole state $\ket{\qh}$ with $N$ particles to the LN state with $N+1$ particles. 
The Rabi-frequency is $\Omega$ and we choose a driving frequency $\omega_c/2$, 
resonant with the zero-interaction energy LN state $\ket{\LN}$. Thereby, in the 
continuum no total angular momentum $\Delta L_z$ is transferred. 

In first order, we couple the quasi-hole state $\ket{\qh}$ and the LN state 
$\ket{\LN}$ to the excitonic states in different particle sectors from $N-1$ to 
$N+2$, as illustrated schematically in Fig. \ref{fig:ap_CoherentPump}. For 
simplicity, only one excitonic state $\ket{N}$ in each particle sector with 
many-body 
gap $\Delta^{(N)}$ is shown. The effective Rabi-frequencies $\Omega_i$, reduced by 
many-body Franck-Condon factors, are labeled by an index $i=1,\ldots,4$.

For the model discussed in section \ref{sec:Continuum}, we find, that the 
Franck-Condon factors 
\begin{align}
 \Omega_2/\Omega &= \langle N | \b_{0,0} | \LN \rangle = 0 \\
\Omega_3/\Omega &= \langle N-1 | \b_{0,0} | \qh \rangle = 0
\end{align}
vanish. Furthermore, we calculate perturbatively the excitation probability 
$P_\Omega$ of excitonic states in first order in $\Omega/\Delta_\LN \ll 1$. Starting 
from the quasi-hole state $\ket{\qh}$, after a $\pi$-pulse of duration $T_\Omega = 
\pi/2\Omega$, we obtain approximately the result in eqn. 
\eqref{eq:FluxInsertionProbability}. There, the factor $\sigma_N$ is defined as
\begin{equation}
 \sigma_N^2 = \L \frac{\pi}{2} \R^2 \sum_j \frac{\Omega_{1 
j}^2/\Omega^2}{\Delta_j^{(N+1)2}/ \Delta_\LN^2} + 
\frac{\Omega_{4 j}^2/\Omega^2}{\Delta_j^{(N+2)2}/ \Delta_\LN^2},
\end{equation}
where the sum over $j$ includes all states in each of the 
particle sectors 
with gap $\Delta_j^{(N)} \geq \Delta_\LN$. For different particle numbers $N$, we 
calculated the prefactor $\sigma_N$ in Fig. \ref{fig:FidelityCoherentPump}.


\end{document}